\documentclass[twocolumn]{aastex63}%
\usepackage{lineno}
\usepackage{amsmath}
\usepackage{mathrsfs}
\usepackage{verbatim}
\usepackage{longtable}
\usepackage{cuted}
\usepackage{appendix}
\usepackage{graphicx}
\usepackage{hyperref}

\submitjournal{ApJL}
\shorttitle{Rebrightening GRB Afterglow in the Stellar Wind Bubble}
\shortauthors{Ren et al.}
\begin{document}
\title{Stellar Wind-Blown Bubbles as Environments for Late-Time Rebrightening of Gamma-Ray Burst Afterglows}

\correspondingauthor{Da-Ming Wei, Jia Ren}
\email{dmwei@pmo.ac.cn}\email{renjia@pmo.ac.cn}

\author[0000-0002-9037-8642]{Jia Ren}
\affiliation{Purple Mountain Observatory, Chinese Academy of Sciences, Nanjing 210034, China}
\author[0000-0002-6588-2652]{Xiao-Yan Li}
\affiliation{Department of Astronomy, Xiamen University, Xiamen, Fujian 361005, China}
\author[0000-0002-8385-7848]{Yun Wang}
\affiliation{Purple Mountain Observatory, Chinese Academy of Sciences, Nanjing 210034, China}
\author[0000-0003-0726-7579]{Lu-Lu Zhang}
\affiliation{College of Physics and Electronic Information Engineering, Guilin University of Technology, Guilin 541004, China}
\author[0000-0002-9758-5476]{Da-Ming Wei}
\affiliation{Purple Mountain Observatory, Chinese Academy of Sciences, Nanjing 210034, China}
\author[0000-0002-7835-8585]{Zi-Gao Dai}
\affiliation{Department of Astronomy, School of Physical Sciences, University of Science and Technology of China, Hefei 230026, China}
\author[0000-0002-9758-5476]{Zhi-Ping Jin}
\affiliation{Purple Mountain Observatory, Chinese Academy of Sciences, Nanjing 210034, China}
\author[0000-0003-1474-293X]{Da-Bin Lin}
\affiliation{Guangxi Key Laboratory for Relativistic Astrophysics, School of Physical Science and Technology, Guangxi University,
Nanning 530004, China}

\begin{abstract}
We presented the multi-wavelength afterglow fitting results for three events that exhibit late afterglow rebrightening behavior: EP240414a ($z=0.401$), GRB 240529A ($z=2.695$), and GRB 240218A ($z=6.782$), which span a broad range of redshifts, from the local to the high-redshift universe.
We prove that the peculiar afterglow light curves of three bursts can be well fitted by structured jets propagated in free-to-shocked stellar wind environment of stellar wind blown bubbles. This scenario offers a self-consistent explanation for the observed subclass of afterglows that exhibit rebrightening that is characterized by steep rises and rapid decays. It also provides a unified solution for such events and offers pathways to study both the jet generation mechanism and the propagation process of jets through the envelope of the progenitor. 
This study reveals that the structured jets produced by such events exhibit a narrow jet core and a steep angle-dependent energy decay index, suggesting highly magnetized jets. The derived transition radii from free stellar winds to shocked stellar winds for all three events are smaller than 0.5~pc, with statistical analysis of similar events indicating a median value of 0.1~pc, which conflicts with numerical simulation results. We anticipate that future observations by EP and SVOM missions will enhance the understanding of analogous events and further reveal information about progenitors and their circum-environments.
\end{abstract}

\keywords
{Gamma-ray bursts (629); High energy astrophysics (739)}
\section{Introduction}
Long gamma-ray bursts (LGRBs) have been observed in association with supernovae~\citep{Galama_Vreeswijk_unusualsupernovaerror_1998}.
Consequently, the progenitors of such events are widely thought to be massive stars at the end stages of their evolution, such as Wolf-Rayet (WR) stars and blue supergiant stars~\citep{Woosley_Bloom_SupernovaGammaRay_2006}. 
Such stars are generally considered to exhibit vigorous convective processes, possess extremely high surface temperatures, drive powerful stellar winds, and may undergo episodic mass ejections. 
In the early 21st century, studies have utilized numerical simulations and observations to investigate the specific structure of the wind-blown bubble surrounding the LGRB progenitor
~\citep[e.g.,][]{RamirezRu_Dray_Windsmassivestars_2001,
Heyl_Perna_BroadbandModelingGRB_2003,
Greiner_Klose_GRB011121Collimated_2003,
Chevalier_Li_DiversityGammaRay_2004,
RamirezRu_GarciaSe_StateCircumstellarMedium_2005,
Eldridge_Genet_circumstellarenvironmentWolf_2006,
Peer_Wijers_SignatureWindReverse_2006,
Chen_Prochaska_AbsenceWindSignatures_2007,
vanMarle_Langer_circumstellarmediumrapidly_2008,
Kong_Wong_Variationmicrophysicswind_2010,
RamirezRu_MacFadyen_HydrodynamicsGammaray_2010}. 
The light curve behavior of LGRB afterglows effectively carries information about the radial evolution of the circumburst environment. This information, in turn, reflects the mass-ejection history of the progenitor during its final evolutionary stages.
It is theoretically expected that the circumburst environment of LGRBs is wind-dominated \citep[e.g.,][and references therein, along with subsequent work]{Eldridge_Genet_circumstellarenvironmentWolf_2006}, a scenario supported by some observational findings \citep[e.g.,][]{Dai_Lu_Gammarayburst_1998,
Fan_Tam_HighenergyEmission_2013,
Piro_Troja_HotCocoonUltralong_2014,
Maity_Chandra_1000DaysLowest_2021,
Varela_Greiner_GRONDgammaray_2025}. 
However, the afterglow behaviors of many LGRBs can also be well explained by a circumburst environment consisting of a homogeneous medium. 
Futhermore, a series of studies based on early afterglow radiation indicates that the circumburst environment may be more complex, deviated from the typical stellar wind or uniform medium environment~\citep[e.g.,][]{Yi_Wu_EarlyAfterglowsGamma_2013,Yi_Wu_BrightReverseShock_2020,Tian_Qin_ConstrainingCircumburstMedium_2022}. 
In other words, the true nature of the circumburst environment remains highly uncertain~\citep[e.g.,][]{Heyl_Perna_BroadbandModelingGRB_2003,
Monfardin_Kobayashi_HighQualityEarly_2006}. 

In the accumulated observational data, there exists a significant number of GRB events exhibiting late-time afterglow rebrightening behavior, which proves difficult to explain using the standard afterglow models~\citep[e.g.,][]{Liang_Li_ComprehensiveStudyGamma_2013}.
During the same period, researchers often utilized the density jumps in the environment to explain the rebrightening phenomenon observed in the GRB afterglows~\citep{RamirezRu_Dray_Windsmassivestars_2001,
Dai_Lu_HydrodynamicsRelativisticBlast_2002,
Lazzati_Rossi_afterglowGRB021004_2002,
Dai_Wu_GRB030226Density_2003,
Tam_Pun_Earlyrebrightenings_2005,
Gendre_Galli_gammarayburst_2007,
Jin_Xu_Xrayafterglow_2009,
Geng_Wu_RevisitingEmissionRelativistic_2014}.
Notably, ~\cite{RamirezRu_Dray_Windsmassivestars_2001} and~\cite{Eldridge_Genet_circumstellarenvironmentWolf_2006} specifically calculated the corresponding afterglow light curves based on the wind-blown bubble simulation results; 
their findings reveal that when the jet propagates into the shocked-wind region, the afterglow displays distinct rebrightening features (however, see~\citealp{Nakar_Granot_Smoothlightcurves_2007,vanEerte_Meliani_Novisibleoptical_2009,vanEerte_Leventis_Gammarayburst_2010,Gat_vanEerte_NoFlaresGamma_2013}).

Recently, the Einstein Probe (EP) satellite detected a series of transient events and their afterglows which identified as fast X-ray transients (FXTs). Owing to rapid multi-wavelength follow-up observations, several FXTs detected by EP have been found to be associated with GRB counterparts. 
In particular, some FXTs that do not have gamma-ray detecting have also detected afterglow-like long-lasting multiband counterparts, e.g., EP240414a and EP241021a.
This suggests the existence of a related population of transients whose origins may be similar to GRBs, but which are distinguished observationally by a lack of detectable gamma-ray emission and prominent X-ray flux
~\citep[e.g.,][]{Gillander_Rhodes_DiscoveryOpticalRadio_2024,
Levan_Jonker_fastXray_2024,
Bright_Carotenut_RadioCounterpartFast_2025,
Srivastav_Chen_IdentificationOpticalCounterpart_2025,
Sun_Li_fastXray_2025,
vanDalen_Levan_EinsteinProbeTransient_2025}.

We note that some of FXT afterglows exhibit late-time rebrightening behavior, e.g., EP240414a~\citep{Sun_Li_fastXray_2025,vanDalen_Levan_EinsteinProbeTransient_2025,Srivastav_Chen_IdentificationOpticalCounterpart_2025}, 
and EP241021a~\citep{Shu_Yang_EP241021aMonthsduration_2025,
Busmann_OConnor_curiouscaseEP241021a_2025}. 
This phenomenon occurs simultaneously in both X-ray and optical bands, showing rapid decay following a brief brightening phase.
We realize that similar phenomena have also been observed past in many GRB afterglows
\citep[e.g.,][]{Laskar_Berger_EnergyInjectionGamma_2015}. 
They were usually interpreted by late energy injection~
\citep[e.g.,][]{Zhang_Meszaros_GammaRayBursts_2002,
Bjoernsso_Gudmundss_EnergyInjectionEpisodes_2004,
Geng_Wu_DelayedEnergyInjection_2013,
Hou_Geng_OriginPlateauLate_2014,
Laskar_Berger_EnergyInjectionGamma_2015,
Yu_Huang_SignatureSpinMagnetar_2015}, refreshed shocks~\citep[e.g.,][]{Rees_Meszaros_RefreshedShocksAfterglow_1998,
Panaitesc_Meszaros_MultiwavelengthAfterglowsGamma_1998,
Granot_Nakar_Astrophysicsrefreshedshocks_2003,
Sun_Geng_GRB240529ATale_2024,
Geng_Hu_GammaRayBurst_2025},
the radiation of $e^\pm$ pairs in the relativistic wind bubble~\citep{Yu_Dai_Shallowdecayphase_2007,
Geng_Wu_ImprintsElectronPositron_2016}, 
density jumps of the medium, 
and two or more component jets~\citep[e.g.,][]
{Huang_Wu_RebrighteningXRF030723_2004,
Guidorzi_Monfardin_EarlyMulticolorAfterglow_2005,
Starling_Wijers_Spectroscopyrayburst_2005,
Filgas_Kruehler_twocomponentjet_2011,
Nardini_Greiner_natureextremelyfast_2011,
Nardini_Elliott_Afterglowrebrighteningsas_2014}.

We observe that the temporal slopes before and after the rebrightening typically differ significantly, with a flatter slope in the pre-phase and a steeper slope in the after-phase. The temporal decay slope after the rebrightening is generally close to the typical slope of a jet-break. The commonality demonstrated by such events remind that they might be driven by the same physical mechanisms and be related to the jet itself.
Spectroscopic observations during rebrightening phases can also provide crucial diagnostics on the nature of the radiating region. 
For instance, the spectroscopic during rebrightening phases of EP240414a is nonthermal dominated ~\citep{Srivastav_Chen_IdentificationOpticalCounterpart_2025,
vanDalen_Levan_EinsteinProbeTransient_2025}, therefore poses a challenge to the pure thermal radiation mechanism model.
The nonthermal models explain for EP240414a include: 
(1) supernova ejecta interacting with the CSM~\citep{vanDalen_Levan_EinsteinProbeTransient_2025}, 
(2) GRB afterglow radiation with some rebrightening effect~\citep{Srivastav_Chen_IdentificationOpticalCounterpart_2025}, 
(3) nonthermal afterglow emission from a mildly relativistic barely failed jet~\citep{Hamidani_Sato_EP240414aGammaRay_2025}, or 
(4) afterglow from a jet with an off-axis viewing angle~\citep{Zheng_Zhu_EP240414aaxisView_2025}.


In this Letter, we attempt to take a holistic view to explain the observational commonalities and attempt to establish a more universal model that is applicable across different redshifts, progenitor star properties, jet structures, and viewing angles. We demonstrate the possibility of such a more unified picture through three distinctive recent events, EP240414a, GRB 240529A, and GRB 240218A. They exhibit the simultaneous late multi-wavelength afterglow brightening behavior. Our research indicates that such events can actually be jointly explained within the framework of wind-blown bubbles, and provides a unified paradigm for similar events.

This paper is organized as follows.
We introduce the acquisition and usage of data in \S~\ref{sec:data}.
In \S~\ref{sec:model}, we introduce our model considerations.
In \S~\ref{sec:analy}, we briefly discuss with analytical derivations how the rebrightening behavior occurs.
In \S~\ref{sec:method}, we introduce our fitting parameters and method in detail.
The results of model parameter inference and corresponding discussion are presented in \S~\ref{sec:discussion}.
We summarize and conclusion the significance of our results in \S~\ref{sec:summary}.
The supporting figures are present in Appendix~\ref{sec:appendix}.
We take the cosmology parameters as
$H_0=67.8~\rm km~s^{-1}~Mpc^{-1}$, and $\Omega_M=0.308$
\citep{PlanckCol_Ade_Planck2015results._2016}.

\section{The acquisition and usage of data}\label{sec:data}
As typical cases, the three events selected in this study, EP240414a ($z=0.401$), GRB 240529A ($z=2.695$), and GRB 240218A ($z=6.782$), span a broad range of redshifts, from the neighboring to the high-redshift universe.
EP240414a is a typical FXT in the neighboring universe that is gamma-ray non-detected with a rest-frame peak energy of the time-integrated spectrum $E_{p,\rm i,z}<1.82$~keV and a very low isotropic energy $E_{X,\rm iso} \sim
5.3_{-0.9}^{+1.2} \times 10^{49}$~erg released in $0.5-4$~keV band \citep{Sun_Li_fastXray_2025}.
GRB 240529A is a typical GRB that has $E_{p,\rm i,z}=595^{+750}_{-370}$~keV and $E_{\gamma,\rm iso}=2.2_{-0.8}^{+0.7}\times 10^{54}$~erg which agree with the Amati relation \citep{Svinkin_Frederiks_KonusWinddetection_2024}.
GRB 240218A is a typical GRB located in high redshift universe that has $E_{p,\rm i,z}=1401^{+599}_{-576}$~keV and $E_{\gamma,\rm iso}=5.4_{-1.6}^{+1.2}\times 10^{53}$~erg which also agree with the Amati relation \citep{Svinkin_Frederiks_KonusWinddetection_2024a}. 

In terms of the afterglow data collecting,
for EP240414a, we collected data from
\cite{Bright_Carotenut_RadioCounterpartFast_2025},
\cite{vanDalen_Levan_EinsteinProbeTransient_2025},
\cite{Srivastav_Chen_IdentificationOpticalCounterpart_2025},
\cite{Sun_Li_fastXray_2025};
for GRB 240529A, we collected data from \emph{Swift} data center, \cite{Sun_Geng_GRB240529ATale_2024} and reference therein;
and for GRB 240218A we use data from \emph{Swift} data center and \cite{Brivio_Campana_comprehensivebroadbandanalysis_2025}.

\section{Model Assumption}\label{sec:model}
Late time achromatic afterglow rebrightening behaviors have been observed in typical GRBs and high-redshift GRB events, for example, 
GRB 060729 ($z=0.54$, \citealp{Grupe_Gronwall_SwiftXMMNewton_2007}), 
GRB 140304A ($z=5.283$, \citealp{Laskar_Berger_VLAStudyHigh_2018}), 
GRB 250221A ($z=0.768$, \citealp{AnguloVal_Becerra_EvidenceEnergyInjection_2025}). See also
\cite{Laskar_Berger_EnergyInjectionGamma_2015} and
\cite{Busmann_OConnor_curiouscaseEP241021a_2025} and reference therein.

The most prominent common feature among these events and the three events in this paper is the presence of a distinct jet-break phase immediately following the rebrightening episodes. 
A successful picture should self-consistently explain the data revealed commonalities.
It is physically natural that a structured jet which produced by the collapse of a massive star has propagate through the wind-blown bubble surrounding the progenitor. 
We propose that the sudden rebrightening observed in the afterglow occurs when the forward shock of the jet passes through the terminal shock region of the stellar wind. 
In the following sections, we will demonstrate this consistency through both analytical and numerical approaches. 
This model provides a self-consistent explanation for the observed subclass of afterglows exhibiting rebrightening.

In this work, we used a state-of-the-art package {\tt ASGARD}\footnote{\url{https://github.com/mikuru1096/ASGARD_GRBAfterglow}}
to generate multi-wavelength afterglow light curves \citep[][see their appendix for details]{Ren_Wang_JetStructureBurst_2024}.
We use numerically solved forward shock dynamics
\citep{Nava_Sironi_Afterglowemissiongamma_2013,
Zhang__physicsgammaray_2018}.
The time-dependent electron continuity equation is solved with first-order accuracy which includes the accurate inverse Compton cooling effect
\citep{Nakar_Ando_Kleinnishinaeffectsoptically_2009}.
We accounted for synchrotron radiation and considered the equal-arrival-time surface effect accordingly~\citep{Waxman__AngularSizeEmission_1997}.

\subsection{Jet Structure}
We consider an axisymmetric structured jet model for bursts and employ a two-part function to describe the shape of the jet,
\begin{equation}\label{eq:jet_str}
	\small
\begin{cases}
	E_{\rm{k,iso}}(\theta) = E_{\rm{k,iso,0}}, 
	\ \Gamma(\theta) = \Gamma_0, 
	& \theta \leqslant \theta_c, \\[1mm]
	E_{\rm{k,iso}}(\theta) = E_{\rm{k,iso,0}} \left( \dfrac{\theta}{\theta_c} \right)^{k_E}, 
	\ \Gamma(\theta) = \Gamma_0 \left( \dfrac{\theta}{\theta_c} \right)^{k_\Gamma}, 
	& \theta_{\rm j} \geqslant \theta > \theta_{\rm c},
\end{cases}
\end{equation}
where $E_{\rm k,iso}(\theta)=4\pi\varepsilon(\theta)$
is the isotropic kinetic energy of jet
with $\varepsilon(\theta)$ being the energy per unit solid angle,
and $\Gamma(\theta)$ is the initial bulk Lorentz factor, respectively.
We artificially set a minimum value of $1.4$ for $\Gamma(\theta)$, corresponding to $\beta \simeq 0.7$. The fitting results presented in this work all meet $\Gamma(\theta_{\rm j})>1.4$.
In many works,
the form of a broken powerlaw function as 
\begin{equation}
	\{E_{\rm k,iso}(\theta),\Gamma_{0}(\theta)\}\propto\left(1+\frac{\theta}{\theta_{\rm c}}\right)^{\{k_E,k_\Gamma\}}
\end{equation}
is adopted.
We point out that this form leads to the strong degeneracy of the core opening angle $\theta_{\rm c}$ and the attenuation exponent $k_{\Gamma}$ and $k_E$ of the jet.
The smooth transition between the core and peripheral regions actually hinders a better limitation of the attenuation index and the viewing angle.

The specific flux calculation is performed based on discretized grids along the ($\theta, \phi$) dimensions. The minimum size ($\delta\theta,\delta\phi$) of the radiation patch is determined by $\delta\theta=\delta\phi \ll 1/\Gamma_0$.

\subsection{Circum-Burst Environment}

For massive stars accompanied by long GRBs, stellar winds are expected to form. The structure of their stellar wind bubble has been studied extensively \citep[e.g.,][]{Castor_McCray_Interstellarbubbles._1975,
	Scalo_Wheeler_PreexistingSuperbubblesas_2001,
	Eldridge_Genet_circumstellarenvironmentWolf_2006}. 
It is generally believed that four regions are formed when stellar winds interact with the interstellar medium (ISM), namely the free stellar wind, the shocked wind, the shocked ISM, and the unshocked ISM \citep{Peer_Wijers_SignatureWindReverse_2006}. 
The number density of the free stellar wind medium is given by $n_f = 3\times10^{35}A_{\star}r^{-2}\text{cm}^{-1}$, where $A_{\star}$ is a constant known as the wind parameter~\citep{Chevalier_Li_WindInteractionModels_2000}.
The number density of the free-to-shocked wind environment is described as
\begin{equation}
	\label{eq:4}
	n(r)=\left\{
	\begin{array}{ll}
		n_f (r), & r < R_{\rm{tr}}, \\
		n_s = \xi n_f (R_{\rm tr}), & r \geqslant R_{\rm tr},
	\end{array}
	\right.
\end{equation}
where $R_{\rm tr}$ is the transition radius of the circum-burst environment from free to shocked wind, and $\xi=\frac{(\hat{\gamma}+1)\mathcal{M}_{\rm wind}^2}{(\hat{\gamma}-1)\mathcal{M}_{\rm wind}^2+2}=4$ is adopted in this paper, where we take $\hat{\gamma}=5/3$ for stellar wind and the Mach number $\mathcal{M}_{\rm wind} \gg 1$.

\section{Analytical Explanation for Afterglow Rebrighting}\label{sec:analy}
Based on~\cite{Li_Lin_LateAfterglowBump/Plateau_2021}, we briefly discuss how the rebrightening behavior occurs.
The derivation is based on the external-forward shock dynamics of a tophat jet with its half-opening angle as $\theta_{\rm j}$.
The specific frequencies $\nu_{\rm m}$ and $\nu_{\rm c}$, and the peak spectral flux $F_{\nu, \max}$ can be calculated.
For on-axis viewer ($\theta_{\rm v}< \theta_{\rm j}$),
\begin{equation}
\left\{
\begin{array}{l}
\nu_{\rm{c}} \propto \bar{\Gamma}_{\rm{obs}}^{8}\left[n\left(\bar{r}_{\rm{obs}}\right)\right]^{3 / 2} \bar{r}_{\rm{obs}}^{-2}, \\
\nu_{\rm{m}} \propto \bar{\Gamma}_{\rm{obs}}^{4}\left[n\left(\bar{r}_{\rm{obs}}\right)\right]^{1 / 2}, \\
F_{\nu, \max} \propto \kappa \bar{\Gamma}_{\rm{obs}}^{2}\left[n\left(\bar{r}_{\rm{obs}}\right)\right]^{3 / 2} \bar{r}_{\rm{obs}}^{3} ;
\end{array}
\right.
\end{equation}
and for off-axis viewer ($\theta_{\rm v}> \theta_{\rm j}+ 1/\bar{\Gamma}$),
\begin{equation}
\left\{
\begin{array}{l}
\nu_{\rm{c}} \propto \bar{\Gamma}_{\rm{obs}}^{6}\left[n\left(\bar{r}_{\rm{obs}}\right)\right]^{3 / 2} \bar{r}_{\rm{obs}}^{-2}, \\
\nu_{\rm{m}} \propto \bar{\Gamma}_{\rm{obs}}^{2}\left[n\left(\bar{r}_{\rm{obs}}\right)\right]^{1 / 2}, \\
F_{\nu, \max } \propto \kappa \bar{\Gamma}_{\rm{obs}}^{-2}\left[n\left(\bar{r}_{\rm{obs}}\right)\right]^{3 / 2} \bar{r}_{\rm{obs}}^{3},
\end{array}
\right.
\end{equation}
where $
\kappa=\left\{1-1 /\left\{1+2 \bar{\Gamma}_{\rm obs}^{2}\left[1-\cos \left(\theta_{\rm j}+\theta_{\rm{v}}\right)\right]\right\}^{2}\right\}$,
$\bar{\Gamma}_{\rm obs}$ and $\bar{r}_{\rm obs}$ are the Lorentz factor and the shock front radius of the jet flow closest to the line of sight.
Besides, the evolution of jet dynamics is
\begin{equation}
\left\{\begin{array}{l}
	\text { for } \bar{r}_{\rm{obs}}<R_{\rm{dec}}: \\ \bar{\Gamma}_{\rm{obs}}=\rm{const}, \bar{r}_{\rm{obs}} \propto t_{\rm{obs}} ; \\ \text { for } \bar{r}_{\rm{obs}}>R_{\rm{dec}}, k=2, \text { and } \theta_{\rm{v}}<\theta_{\rm{jet}}: \\ \bar{\Gamma}_{\rm{obs}} \propto t_{\rm{obs}}^{-1 / 4}, \bar{r}_{\rm{obs}} \propto t_{\rm{obs}}^{1 / 2} ; \\ \text { for } \bar{r}_{\rm{obs}}>R_{\rm{dec}}, k=2, \text { and } \theta_{\rm{v}}>\theta_{\rm{jet}}+1 / \bar{\Gamma}: \\ \bar{\Gamma}_{\rm{obs}} \propto t_{\rm{obs}}^{-1 / 2}, \bar{r}_{\rm{obs}} \propto t_{\rm{obs}} ; \\ \text { for } \bar{r}_{\rm{obs}}>R_{\rm{dec}}, k=0, \text { and } \theta_{\rm{v}}<\theta_{\rm{jet}}: \\ \bar{\Gamma}_{\rm{obs}} \propto t_{\rm{obs}}^{-3 / 8}, \bar{r}_{\rm{obs}} \propto t_{\rm{obs}}^{1 / 4} ; \\ \text { for } \bar{r}_{\rm{obs}}>R_{\rm{dec}}, s=0, \text { and } \theta_{\rm{v}}>\theta_{\rm{jet}}+1 / \bar{\Gamma}: \\ \bar{\Gamma}_{\rm{obs}} \propto t_{\rm{obs}}^{-3 / 2}, \bar{r}_{\rm{obs}} \propto t_{\rm{obs}} ,
\end{array}\right.
\end{equation}
where $R_{\rm dec}$ is the deceleration radius of the jet, and $k=2$ corresponding to free wind medium,
$k=0$ for uniform medium, respectively.
Based on the results of~\cite{Sari_Piran_SpectraLightCurves_1998}, and adopting the observed spectral flux
$f_{\nu}\left(t_{\rm{obs}}\right) \propto t_{\rm{obs}}^{-\alpha} \nu^{-\beta}$,
then the closure relations between $\alpha$ and $\beta$ can be derived \citep[e.g.,][]{Zhang_Meszaros_GammaRayBursts_2004,
Zhang_Liang_ComprehensiveAnalysisSwift_2007}.

For the optical afterglow, the typical scenario considers the $\nu_{\rm m} < \nu_{\rm opt} <\nu_{\rm{c}}$ under slow cooling conditions.
If the jet is in the coasting phase while sequentially traversing both the free stellar wind and the uniform medium, i.e., $\bar{r}_{\rm{obs}}<R_{\rm{dec}}$, then one can have $\alpha=(p-1)/2$ (wind) and $\alpha=-3$ (uniform), where $p\in (2,3)$ is the powerlaw index of shock-accelerated nonthermal electrons.  It is effective both for on-axis and off-axis scenario.
Another possible scenario is when the jet is decelerated, $\bar{r}_{\rm{obs}}>R_{\rm{dec}}$, and is observed off-axis-viewed.
One can derive $\alpha=p-2$ (wind) and $\alpha=-(15-3p)/2$ (uniform).
In both scenarios, the optical afterglow can exhibit a transition from gradual decay to rapid brightening, naturally explaining the rebrightening slopes ranging from $t^3$ to $t^{4.5}$. This aligns remarkably well with the observational data from the two EP events EP240414a ($t^3$, \citealp{Zheng_Zhu_EP240414aaxisView_2025}) and EP241021a ($t^4$, \citealp{Busmann_OConnor_curiouscaseEP241021a_2025}).
There are other scenarios that can produce rebrightening behavior, please see~\cite{Li_Lin_LateAfterglowBump/Plateau_2021} for more details.

The transition of circum-jet environment from the free wind region to the shocked wind region does not necessarily involve a complete deceleration of the jet.
This implies that after the rebrightening, the afterglow may still maintain a slope of normal decay for a period of time, as observed in EP241021a.
For events like EP240414a, the model needs that the Lorentz factor of the jet during rebrightening coincides with the opening angle of the jet core $1/\Gamma(t_{\rm obs})\sim \theta_{\rm c}$, thereby producing the peculiar steep decay behavior, namely a ``jet-break".
When other parameters are held constant, variations in the viewing angle and jet structure can naturally shift the afterglow light curve from classical behavior to a late rebrightening pattern.

\section{Numerical Fitting and Results}\label{sec:method}

\subsection{Fitting Method}

\begin{deluxetable*}{lcrrr}[htbp]
\label{tab1}
\tablecaption{\textbf{Fitting Results}}
\tablehead{
\colhead{Parameter} & \colhead{Range\tablenotemark{a}} & & \colhead{Value\tablenotemark{b}} &  \\\hline & & \colhead{EP240414a} & \colhead{GRB 240529A} & \colhead{GRB 240218A}}
\startdata
$\log_{10}n_s$ (cm$^{-3}$) & $[-2,1]$  & $0.65^{+0.24}_{-0.36}$ & $0.06^{+0.07}_{-0.09}$ & $0.52^{+0.27}_{-0.28}$ \\
$\log_{10}A_{\star}$       & $[-2,1]$ & $0.46^{+0.25}_{-0.36}$ & $-0.10^{+0.05}_{-0.06}$ &  $-0.52^{+0.21}_{-0.17}$ \\
$\log_{10}E_{\rm k,iso}$ (erg) & $[53.5,56.5]$  & $54.97^{+0.24}_{-0.35}$  & $56.06^{+0.05}_{-0.05}$  & $55.21^{+0.17}_{-0.11}$  \\
$p$             & $[2,3]$   &  $2.13^{+0.02}_{-0.042}$  &  $2.05^{+0.00}_{-0.02}$ &  $2.08^{+0.03}_{-0.03}$ \\
$\log_{10}\Gamma_{0}$    & $[1.5,3]$  & $1.79^{+0.20}_{-0.13}$  & $2.13^{+0.01}_{-0.01}$ & $2.50^{+0.07}_{-0.10}$ \\
$\log_{10}\epsilon_{e}$  & $[-3,-0.5]$   & $-2.30^{+0.35}_{-0.25}$  & $-0.77^{+0.03}_{-0.04}$ & $-0.99^{+0.13}_{-0.19}$ \\
$\log_{10}\epsilon_{B}$  & $[-6,-0.5]$   & $-2.65^{+0.37}_{-0.25}$  & $-2.20^{+0.10}_{-0.08}$ & $-3.44^{+0.26}_{-0.29}$ \\
$\log_{10}\xi_{e}$  & $[-3, 0]$   & $-1.45^{+0.36}_{-0.25}$  & $-0.02^{+0.01}_{-0.02}$ & $-0.21^{+0.13}_{-0.18}$ \\
$\log_{10}\theta_{\rm c}$ (rad)       & $[-3,0]$   & $-1.78^{+0.03}_{-0.03}$  & $-2.60^{+0.03}_{-0.03}$ & $-1.92^{+0.05}_{-0.06}$ \\
$\log_{10}(\theta_{\rm j}/\theta_{\rm c})$  & $[0,1]$  & $0.34^{+0.06}_{-0.05}$  & $0.64^{+0.03}_{-0.06}$ & $0.57^{+0.26}_{-0.10}$ \\
$\theta_{\rm v}/\theta_{\rm c}$  & $[0,10]$   & $6.36^{+0.34}_{-0.29}$  & $3.54^{+0.21}_{-0.23}$ & $2.36^{+0.28}_{-0.18}$ \\
$k_{\Gamma}$  & $[-8,0]$   & $-1.82^{+1.18}_{-0.82}$  & $-2.12^{+0.24}_{-0.29}$ & $-0.21^{+0.16}_{-0.23}$\\
$k_E$  & $[-10,0]$   & $-8.60^{+0.70}_{-0.77}$  & $-2.56^{+0.18}_{-0.18}$ & $-4.40^{+0.51}_{-0.64}$ \\
$E(B-V)_{\rm host}$  & $[0,1]$   & 0(fixed)  & $0.15^{+0.02}_{-0.02}$ & $0.04^{+0.01}_{-0.01}$ \\
$R_{V, \rm host}$  & $[1,5]$   & --  & $1.32^{+0.30}_{-0.23}$ & $2.84^{+0.73}_{-0.58}$ \\
\hline
$R_{\rm tr}$   (cm)     & -- & $8.84\times 10^{17}$  & $9.15\times 10^{17}$ & $3.32\times 10^{17}$ \\
\enddata
\tablenotetext{a}{Uniform prior distribution.}
\tablenotetext{b}{Errors in $1\sigma$.}
\end{deluxetable*}

\begin{figure}[htbp]
	\centering
	\includegraphics[width=0.42\textwidth]{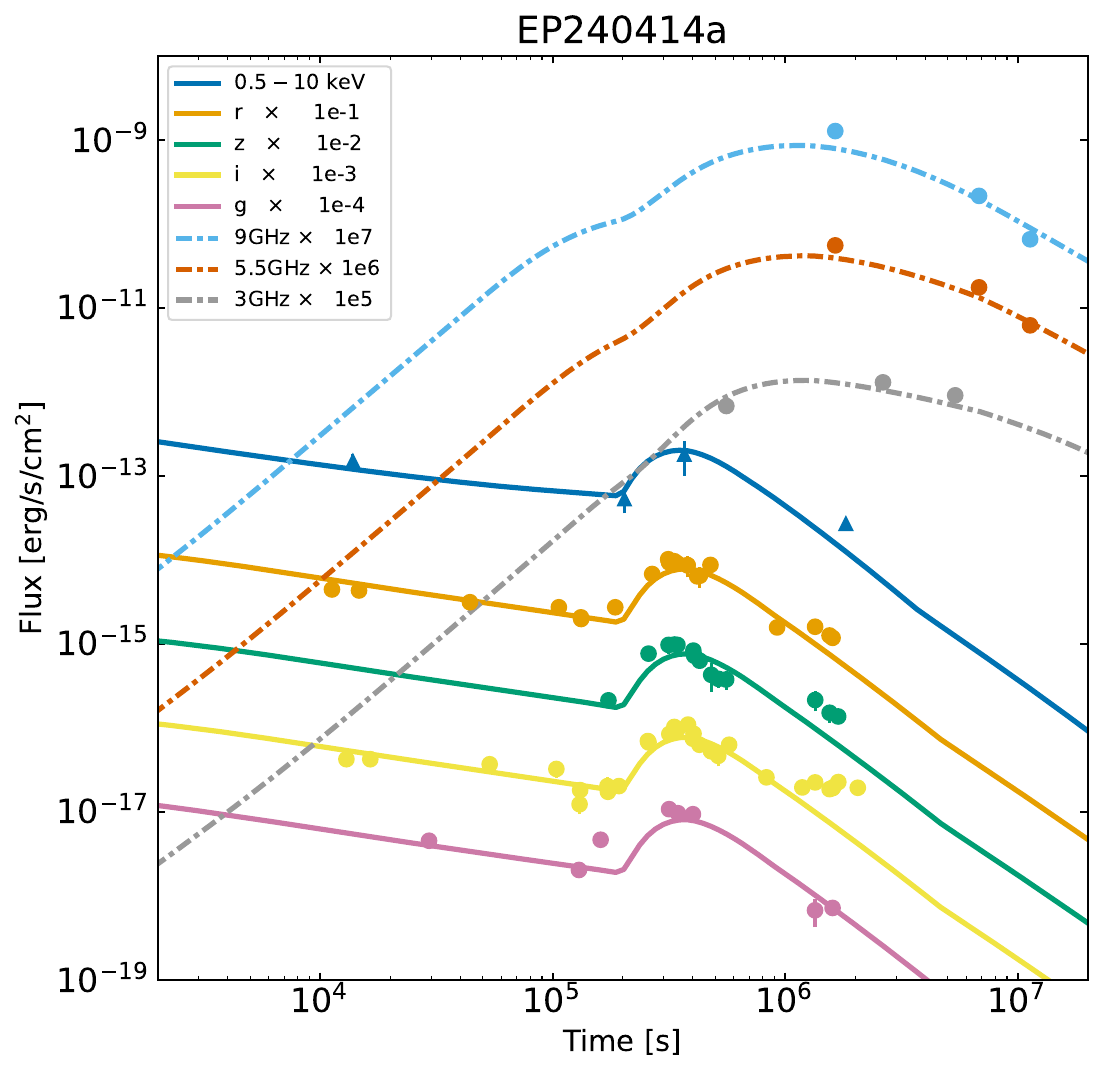}\\
	\includegraphics[width=0.42\textwidth]{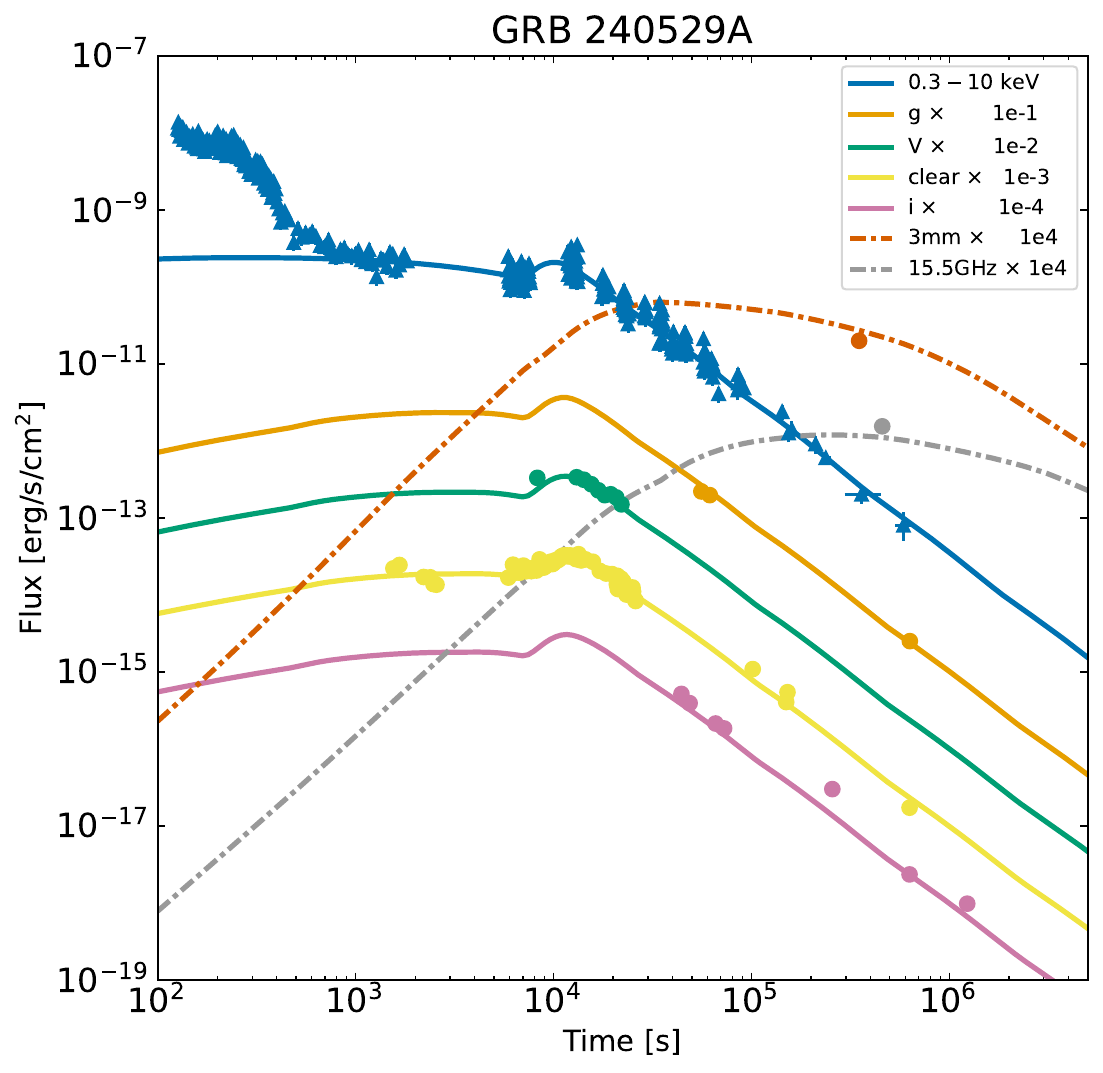}\\
	\includegraphics[width=0.42\textwidth]{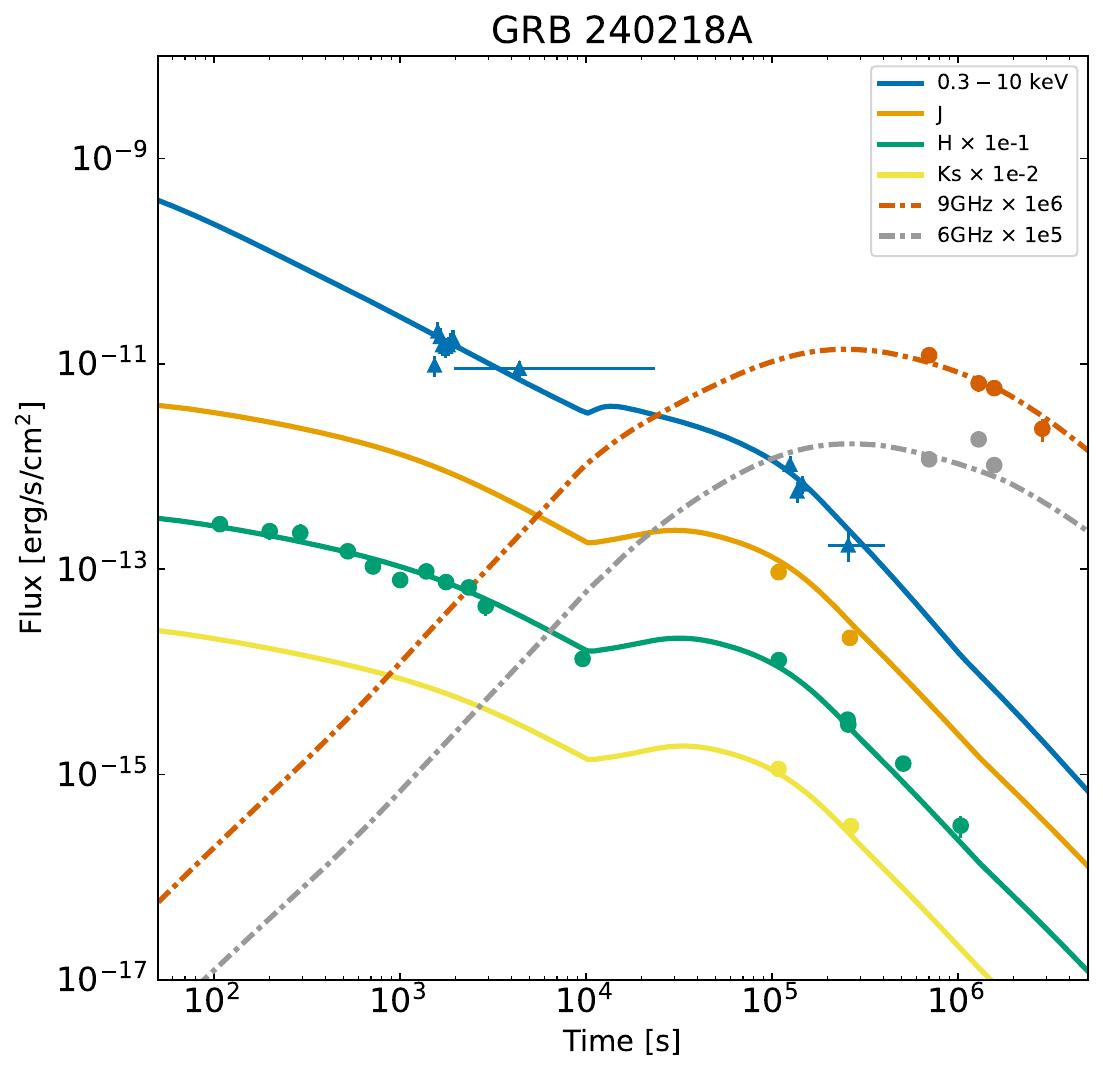}
	\caption{The best-fit light curves for the three events. From top to bottom: a local low-luminosity burst EP240414a, a typical nearby GRB 240529A, and a high-redshift GRB 240218A, respectively.}.
	\label{LCs}
\end{figure}

We perform posterior parameter inference using
the Markov Chain Monte Carlo (MCMC) technique,
employing the Python package {\tt emcee} as the sampler
\citep{ForemanMa_Hogg_emceeMCMCHammer_2013}.
The sampling was carried out in multi-dimensional parameter space.
The basic model parameters include
\{$n_s$, $A_{\star}$, $E_{\rm k,iso}$, $p$, $\Gamma_{0}$, 
$\epsilon_{e}$, $\epsilon_{B}$, $\xi_{e}$,
$\theta_{\rm c}$, $\theta_{\rm j}$,
$\theta_{\rm v}$, $k_{\Gamma}$, $k_{E}$\},
where $\epsilon_{e}$, $\epsilon_{B}$, $\xi_{e}$
are the energy fraction of electrons and magnetic field in the jet and the number fraction of nonthermal electrons to the total shock swept electrons, respectively.
We also assume the SMC-like extinction template for host galaxies \citep{Pei__Interstellardustmilky_1992},
and setting $E(B-V)_{\rm host}$ and $R_{V, \rm host}$ as fitting parameters.
According to \cite{Sun_Li_fastXray_2025}, we ignored the host galaxy extinction of EP240414a, and for GRB 240218A and GRB 240529A, both parameters are inferred by the fitting.
The best-fit lightcurves of these bursts are presented in Figure~\ref{LCs}.
The corresponding parameter values are detailed in Table~\ref{tab1}.
After discarding the burn-in stage, the resulting marginalized posterior probability density distributions are presented in Appendix~\ref{sec:appendix}, Figure~\ref{corner_EP240414a} for EP240414a, Figure~\ref{corner_240529A} for GRB 240529A, and Figure~\ref{corner_240218A} for GRB 240218A, respectively.

\subsection{Basic results}

We find the wind parameter $A_{\star}$ are $\sim 2.9$, $\sim 0.8$, and $\sim 0.3$ of EP240414a, GRB 240529A, and GRB 240128A, respectively. Thus they are consistent with the WR progenitor scenario. The kinetic energy of these bursts are located in the high level of energy distribution of GRBs. 
For the two typical GRBs, GRB 240529A has isotropic-equivalent prompt emission energy 
$E_{\gamma,\rm iso}=2.2_{-0.8}^{+0.7}\times 10^{54}$~erg
~\citep{Svinkin_Frederiks_KonusWinddetection_2024}, 
in turn, the prompt emission efficiency is $\eta_\gamma=E_{\gamma,\rm iso}/(E_{\gamma,\rm iso}+E_{\rm k,iso}) \sim 1.9$\%; 
GRB 240218A has $E_{\gamma,\rm iso}\sim 3.24\times 10^{53}$~erg
~\citep{Brivio_Campana_comprehensivebroadbandanalysis_2025}, 
thus $\eta_\gamma \sim 2.0$\%. 
We note that in our fittings,  both bursts are off-axis-viewed (GRB 240529A, $\theta_{\rm v}/\theta_{\rm c} \sim 3.54$; GRB 240218A, $\theta_{\rm v}/\theta_{\rm c} \sim 2.36$). 
Thus the relatively low radiation efficiencies can be understood.
Specially, EP240414a has a very low $E_{X,\rm iso} \sim
5.3_{-0.9}^{+1.2} \times 10^{49}$~erg in $0.5-4$~keV band and was lack of gamma-ray emission
~\citep{Sun_Li_fastXray_2025}.
Although the fitting yields an isotropic-equivalent kinetic energy $E_{\rm k,iso}\sim 9.3 \times10^{54}$~erg, this scenario remains highly self-consistent due to the extremely large off-axis viewing angle ($\theta_{\rm v}/\theta_{\rm c} \sim 6.36$).

The bulk Lorentz factors of these bursts are typical, the median distribution value is $\Gamma_0 \sim 62$ of EP240414a, $\Gamma_0 \sim 135$ of GRB 240529A, and $\Gamma_0 \sim 316$ of GRB 240218A, respectively.
For all three bursts we observe very narrow core opening angle, $\theta_{\rm c} \sim 0^{\circ}.95$ of EP240414a, $\theta_{\rm c} \sim 0^{\circ}.14$ of GRB 240529A, and $\theta_{\rm c} \sim 0^{\circ}.69$ of GRB 240218A, respectively.  We will give possible explanation in \S~\ref{sec:JetStructure} in detail.
We get typical microphysical parameters of them.

\subsection{Transition Radius}

The transition radius, which indicates where the inner radius of the terminal shock of the wind is, can be calculated with $n_s = \xi n_f(R_{\rm tr}) = 3 \times 10^{35} A_{\star} R_{\rm tr}^{-2} \xi$. Thus $R_{\rm tr}=1.1 \times 10^{18}\; A_{\star}^{1/2} n_s^{-1/2}$~cm.
The best-fit results of $R_{\rm tr}$ are listed in Table~\ref{tab:comparison}, we find all three bursts have small transition radius $<0.5$~pc.
We also collected some bursts that were considered to involve free-to-shocked wind conversion behaviors and listed them in Table~\ref{tab:comparison}.
One can observe all sample values are less than 1~pc.
By fitting those samples with a log-normal function, we found that the median value is ${\rm log_{10}}R_{\rm tr}{(\rm cm)}=17.59$, with a standard deviation STD=0.45, as shown in Figure~\ref{Radii}.

Based on the different progenitor stars and the properties of the burst environment, recently \cite{Chrimes_Gompertz_Towardsunderstandinglong_2022}  conducted a systematic study on the characteristics of the stellar wind bubbles around LGRBs through a series of numerical simulations. Their statistic results indicate that the median value of the wind termination shock radius is located at 10~pc.
Thus, the above results indicate that the inferred conversion radius based on observations shows a systematic deviation from the numerical simulation results. 
The exact cause of this deviation remains unknown.
However, we hypothesize that one possible explanation is that the GRB progenitors are extremely unstable and collapsed during the early stage of rapid mass loss process, preventing the stellar wind bubble from expanding to the size predicted by the numerical simulation
~\cite[e.g.,][]{RamirezRu_GarciaSe_StateCircumstellarMedium_2005}.
Besides, the simulation results are taken place in one dimensional, 
the instability that only occurs in high-dimensional situations, e.g., Rayleigh-Taylor and Kelvin-Helmholtz Instabilities, may lead to a smaller transition radius.

\begin{figure}[htbp]
	\centering
	\includegraphics[width=0.45\textwidth]{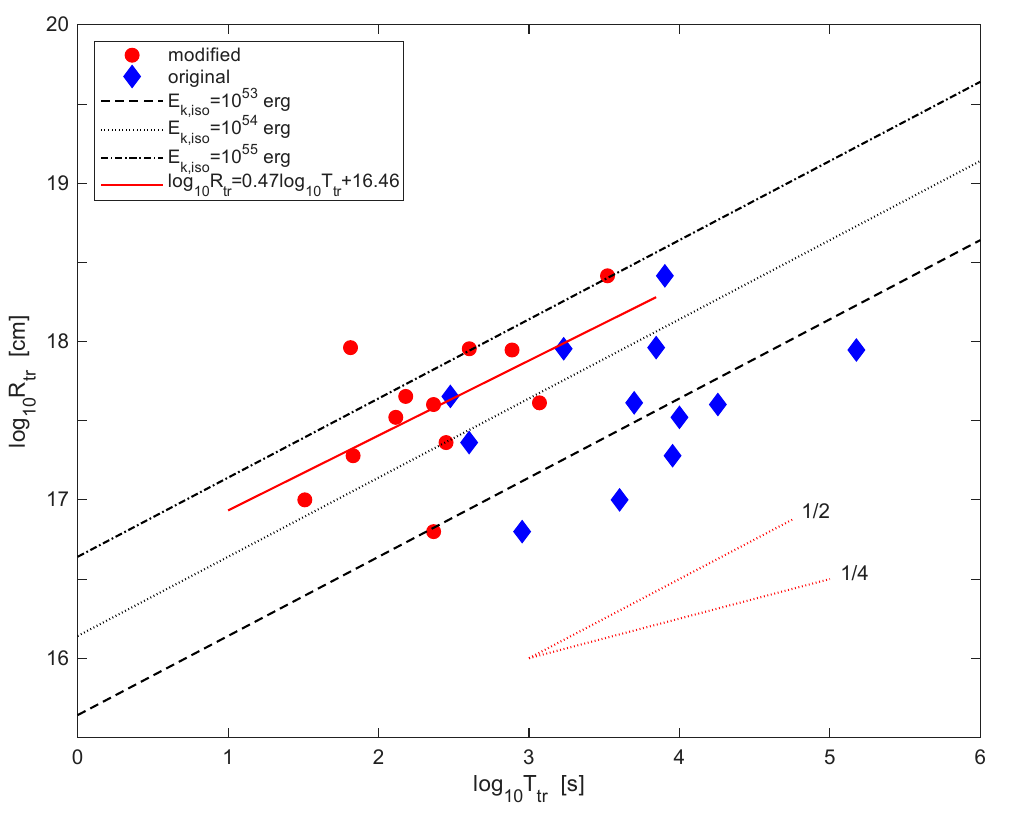}
	\caption{The correlation between the transition radii $R_{\rm tr}$ and the transition times $T_{\rm tr}$. The blue diamonds represent the observed values without considering redshift and viewing angle corrections from Equation~\ref{T_tr}, while the red dots show the corrected results. 
	The red solid line presents a loglinear fit of corrected results.
	The dashed, solid, and dash-dotted lines correspond to the relationships between jet propagation distance and time under different jet kinetic energies, respectively. The results are calculated according to Equation~\ref{R_wind}, where we have taken $A_\star = 0.3$. 
    We also present reference lines with slopes of 1/2 and 1/4, respectively.}
	\label{Distribution}
\end{figure}

In Figure~\ref{Distribution}, we present the correlation between the transition radii $R_{\rm tr}$ and the transition times $T_{\rm tr}$. 
We define $T_{\rm tr}$ as the onset moment when the broadband light curve exhibits rebrightening behavior.
The specific values are also listed in the Table~\ref{tab:comparison}.
For the convenience of analysis, we also provide the redshift and off-axis factor $\theta_{\rm v}/\theta_{\rm c}$ of each event.
We point out that since the rebrightening takes place in a uniform medium environment (see Section~\ref{sec:analy}), the occurrence time of rebrightening can be described by the following relation \citep[e.g.,][Equation~31]{Beniamini_Granot_Afterglowlightcurves_2020},
\begin{equation}\label{T_tr}
	T_{\rm tr} 
	\propto \left( \frac{\theta_{\rm v}}{\theta_{\rm c}}\right)^\frac{2(4-k)}{3-k}
	=
	\left\{
	\begin{array}{lr}
	\left( \theta_{\rm v}/\theta_{\rm c}\right)^{8/3}, & \theta_{\rm v} \ge \theta_{\rm c}; \\
	1,  & \theta_{\rm v} < \theta_{\rm c}.
	\end{array}
	\right.
\end{equation}
After taking into account the redshift correction, we apply the above relationship to convert the transition time into the on-axis situation.
We also plot reference lines to characterize the relationship between the propagation distance of GRB jets in a wind-like medium and time~\citep{Granot_Sari_ShapeSpectralBreaks_2002},
\begin{equation}\label{R_wind}
	R=\left[ \frac{(17-4k)(4-k)E_{\rm k,iso}t^{\prime}}{4\pi A c}\right]^{1/(4-k)}, 
\end{equation}
where $A=5\times 10^{11} {\rm g\;cm^{-1}} A_{\star}$, and take $k=2$.
Basing on a log-linear function fit we find the corrected samples follows the relation as 
\begin{equation}
	{\rm log}_{10}R_{\rm tr}({\rm cm})=0.47{\rm log}_{10}T_{\rm tr}+16.46.
\end{equation}
It can be clearly seen in Figure~\ref{Distribution} that samples follow the 1/2 slope expected for a wind environment, rather than the 1/4 slope characteristic of a homogeneous environment. 

\section{Discussion}\label{sec:discussion}

\begin{deluxetable*}{crrrrc}[htpb]
	\label{tab:comparison}
	\tablecaption{\textbf{Possible Free-to-Shocked Wind Environment Transients}}
	\tablehead{
		\colhead{Transients} & \colhead{Transition Radius} & \colhead{Transition Time} & \colhead{$\theta_{\rm v}/\theta_{\rm c}$} & \colhead{redshift}
		& \colhead{Reference} \\
		\colhead{} & \colhead{($\times 10^{17}$cm)} & \colhead{(s)} & \colhead{} &	\colhead{} & \colhead{}}
	\startdata
	GRB 050319  & $(3.0-15.0)$  & $1700$ & 0 & 3.24 & \cite{Kamble_Resmi_ObservationsOpticalAfterglow_2007} \\
	GRB 081109A  & $ 4.5$  & $300$ & 0 & 0.98 & \cite{Jin_Xu_Xrayafterglow_2009} \\
	GRB 100814A & $ 4.0$  & $18000$ & 3.66 & 1.44 & \cite{Li_Lin_LateAfterglowBump/Plateau_2021} \\
	GRB 100901A & $ 1.9 $  & $9000$ & 4.50 & 1.408 & \cite{Li_Lin_LateAfterglowBump/Plateau_2021} \\
	GRB 120326A & $1.0 $  & $4000$ & 4.14 & 1.798 & \cite{Li_Lin_LateAfterglowBump/Plateau_2021} \\
	GRB 120404A & $ 0.63$  & $900$ & 0.68 & 2.876 & \cite{Li_Lin_LateAfterglowBump/Plateau_2021} \\
	GRB 140423A  & $ 4.1$  & $5000$ & 0 & 3.26 & \cite{Li_Wang_GRB140423ACase_2020} \\
	GRB 160625B  & $ 26.0$  & $8000$ & 0 & 1.406 & \cite{Fraija_Veres_TheoreticalDescriptionGRB_2017} \\
	GRB 190114C  & $ 2.3$  & $400$ & 0 & 0.424 & \cite{Fraija_Dichiara_AnalysisModelingMulti_2019} \\
	EP240414a  & $ 8.8$  & $150000$ & 6.36 & 0.402 & this work \\
	GRB 240529A  & $ 9.2$  & $7000$ & 3.54 & 2.695 & this work \\
	GRB 240218A  & $ 3.3$  & $10000$ & 2.36 & 6.782 & this work \\
	\enddata
\end{deluxetable*}

\subsection{Jet Structure}
\label{sec:JetStructure}

Very narrow jet cores with $\theta_{\rm c} \lesssim 1^{\circ}$ have been obtained for all three bursts in this work. In the model adopted in this paper, narrower jets facilitate the production of more sharper rebrightening features, transitioning directly from the peak flux phase to the jet-break stage, as demonstrated in Figure~3 of \cite{Li_Lin_LateAfterglowBump/Plateau_2021}.
Besides,  \cite{Li_Lin_LateAfterglowBump/Plateau_2021} also suggest small $\theta_{\rm c}$ of GRBs 120326A, 100901A, 100814A, and 120404A. 
Although their structured jet model was a Gaussian jet, we suppose both results have comparability due to the very sharp jet edge from this work. 
Such a small $\theta_{\rm c}$ requires the core region of the jet to be highly collimated. However, there has been little systematic research on the generation mechanisms of extremely narrow jets, for example, through numerical simulations.
Fortunately, the brightest-of-all-time GRB 221009A provides key evidence for the existence of a highly collimated jet. 
A series of works have inferred that the narrow jet opening angle of GRB 221009A is $\lesssim 1^{\circ}$
\citep[e.g.,][]{LHAASOCol_Cao_teraelectronvolt_2023,
Ren_Wang_JetStructureBurst_2024,
Zhang_Murase_originveryhigh_2025}.
The fitting results in this paper thus gain crucial observational support.
Further, narrow jets may not be uncommon in GRBs. 
Some bright X-ray flares comparable to or even stronger than the prompt emission~\citep[e.g.,][]{Falcone_Burrows_GiantXRay_2006} could be emitted by narrow jets slightly offset from the line of sight. 
Observers would miss the prompt emission of narrow jet entirely in most cases. 
Under this scenario, most GRBs are observed within the wing of a structured jet~\citep{Zhang_Wang_BOATGRB221009A_2024}.

The relativistic jets that successfully break out from the progenitor's envelope can indeed produce very narrow jet opening angles. The post-breakout jet opening angle, $\theta_{\rm j}$, depends on the initial jet opening angle $\theta_0$ injected by the central engine into the progenitor's envelope, the jet power, and the progenitor's radius and density profile. For relativistic hydrodynamical jet, one can find 
$\theta_{\rm j}/\theta_0 \sim 1/10$ can be easy to achieve~\citep[][Equation 35]{Hamidani_Ioka_Jetpropagationexpanding_2021}.
However, the realistic value of $\theta_0$ is uncertain, and it is often assumed in literatures that $\theta_0\sim 10^\circ$, so that $\theta_{\rm j}\sim 1^\circ$ is natural. 
For magnetized jets, the powerful pinch effect from well-ordered magnetic fields helps protect the relativistic jet from severe baryon contamination by the cocoon material and confines the jet to smaller opening angles \citep{Gottlieb_Bromberg_structureweaklymagnetized_2020}. 
Consequently, magnetically dominated jets are more promising to possess narrower jet cores.

Basing on the fitting results, we have the angular distribution indexes of jet as $k_{\Gamma} = -1.82^{+1.18}_{-0.82}$,  $-2.12^{+0.24}_{-0.29}$, $-0.21^{+0.16}_{-0.23}$;
$k_E = -8.60^{+0.70}_{-0.77}$,  $-2.56^{+0.18}_{-0.18}$, $-4.40^{+0.51}_{-0.64}$,
for EP240414a, GRB 240529A, and GRB 240218A, respectively. 
Noticeably , the $k_{\Gamma}$ value of GRB 240218A tends toward the upper limit of the parameter range.
The lack of observational data during the rising phase of the rebrightening stage may resulting in poor constraints on the contribution from the large-angle radiation regions of the jet. 
Besides, the slow part of jet in the wing region can modulate the slope of the rising phase by increasing the contribution of radiation from earlier deceleration processes.
Thus, the data lacking leads to a poor restriction on the Lorentz factor profile.
All bursts possess highly relativistic jets and exhibit a pronounced attenuation of the wing component in their energy distribution, which aligns well with the results of numerical simulations.

$k_{\Gamma}$ and $k_E$ exhibit significant variations across different bursts. 
It has suggested that the distribution of baryon loading may be a more fundamental physical quantity~\citep{Ren_Lin_ConstrainingJetLaunching_2020}.
Because $\Gamma(\theta) \propto E_{\rm k,iso}(\theta)/M_{\rm jet}(\theta) c^2$, thus the angle-dependent mass distribution index $k_M=k_E-k_\Gamma$ characterizes the distribution of baryon loading in the jet. It is easy to derive that $k_M \sim -6.4, -0.4, -4.2$ for for EP240414a, GRB 240529A, and GRB 240218A, respectively.
 Baryon loading characterizes the mixing intensity of matter at the interface within the jet-cocoon system and from the jet head region, potentially hinting at the mechanisms of jet-cocoon interactions within collapsar.

In relativistic magnetohydrodynamic (RMHD) simulations of jets, the energy and Lorentz factor often exhibit steep decay slopes related to $\theta$ when beyond a critical angle $\theta_{\rm c}$.
For example, \cite{Tchekhovs_McKinney_Simulationsultrarelativisticmagnetodynamic_2008} studied RMHD jets and showed that $k_E < -4$ and could even reach down to $-10$; $k_\Gamma$ may have a two-segment structure ranging from gentle to steep, transitioning from $-1$ to $-2.5$. 
What needs illustration is that the entrainment and instabilities within the jet may cause the distribution to become less sharp at large angles.
Overall, the fitting results presented in this work are consistent with the RMHD jet simulations. 


\subsection{Reasons for the non-Detection of EP240414a Gamma-rays}

Here we use the methods described in \cite{Ioka_Nakamura_Spectralpuzzleaxis_2019} to calculate the isotropic energy of prompt emission 
\begin{equation}
	E_{\gamma, \rm iso }=\int \frac{\sin \theta \mathrm{d} \theta}{2} E_{\gamma}(\theta) \cdot \mathcal{B}(\theta),
\end{equation}
where the beaming term is
\begin{equation}
	\mathcal{B}(\theta) = \frac{2\bigl[1 - \beta(\theta)\cos\theta\cos\theta_{\rm v}\bigr]^{2} + \bigl[\beta(\theta)\sin\theta\sin\theta_{\rm v}\bigr]^{2}} {2\Gamma^{4}(\theta)\Bigl\{\bigl[1 - \beta(\theta)\cos(\theta_{\rm v}+\theta)\bigr]\bigl[1 - \beta(\theta)\cos(\theta_{\rm v}-\theta)\bigr]\Bigr\}^{5/2}}. 
\end{equation}
Here $E_{\gamma}(\theta)=\epsilon_\gamma E_{\rm k,iso}(\theta)$, and $\epsilon_\gamma$ being the radiation efficiency of prompt emission.
When we incorporate the structured jet parameters fitted in this paper, 
we obtain $E_{\gamma, \rm iso}/\epsilon_\gamma = 7.9\times 10^{50}$~erg.
If we take $E_{\gamma, \rm iso}=E_{X, \rm iso}=5.3 \times 10^{49}$~erg of EP240414a, the corresponding value of $\epsilon_\gamma \sim 0.07$ is quite reasonable.
The on-axis-viewed isotropic prompt emission energy then be $E_{\gamma, \rm iso, on} \sim 5.1 \times 10^{53}$~erg, which is about 4 orders of magnitude higher than the derived value from realistic observations.

Since the jet Lorentz factor and energy decrease rapidly out of the jet core, we are approximately derive the prompt spectral peak energy for on-axis observation using the analytical relationships in the top-hat jet model
\citep{Granot_Panaitesc_AxisAfterglowEmission_2002,
Granot_Guetta_LessonsShortGRB_2017},
\begin{equation}
	E_{p,\rm on}=\bigl[1+\Gamma_{0}(\theta_{\rm{v}}-\theta_{\rm c})^2\bigr] E_{p,\rm obs}.
\end{equation}
From this, it is calculated that $E_{p,\rm on}(1+z)<56.6$~keV.
On the well-known Amati relation plane, the corrected value of EP240414a, which we have inferred for off-axis observation, lies out of the $2\sigma$ lower boundary of the Type II GRB category and maybe similar to the supernova-associated high luminosity GRBs
\citep{Sun_Li_fastXray_2025,Wang_Ren_RapidGBMEfficientTool_2025}.
Based on the calculation results from this section and referring to extended data Figure~1 in \cite{Sun_Li_fastXray_2025}, we find that the corrected instantaneous radiation spectrum of EP240414a naturally falls above the observational flux thresholds provided by Swift-BAT and Konus-Wind in the $15–350$~keV energy range. 
Therefore, the model self-consistently explains the lack of gamma-ray observations for this event as it is an off-axis-viewed GRB. 
Our calculation results suggest a possible connection between off-axis observed GRBs and some FXTs, and indicate that the large $\theta_{\rm{v}}/\theta_{\rm c}$ ratio may be a key to understanding such FXTs.

\subsection{Fomation Rate}

\cite{Rastineja_Levan_EP250108a/SN2025kg_2025} estimated the occurrence rate of FXT events is at least few$\times 10$~Gpc$^{-3}$ yr$^{-1}$, 
which is higher than the typical high luminosity GRBs rate of $\sim1$~Gpc$^{-3}$ yr$^{-1}$ \citep{Sun_Zhang_ExtragalacticHighenergy_2015}. 
\cite{Sun_Li_fastXray_2025} also inferred a local event rate density similar to EP240414a as $< 1$~Gpc$^{-3}$ yr$^{-1}$, and upper limit for FXT events without gamma-ray counterparts as $< 10$~Gpc$^{-3}$ yr$^{-1}$.
Based on the model presented in this paper, it can be naturally attributed to the off-axis observation of the jet. 
On the one hand, based on the fitting result, the EP240414a event can still be observed by EP when $\theta_{\rm v} > 6 \theta_{\rm c}$. 
From this, it can be roughly estimated that the observation rate of similar events can reach several tens of times that of typical GRB events, matching the results inferred by \cite{Rastineja_Levan_EP250108a/SN2025kg_2025}.
Under the scenario of off-axis jet observations, FXT events without gamma-ray counterparts correspond to cases with extremely large viewing angles, representing a subclass of off-axis-viewed events. Meanwhile, the detection is also constrained by instrumental selection effects, hence the actual event rate should be lower than our rough estimate.
On the other hand, we constrained the energy decay slope of the structured jet's power-law wing to be more sharp than $\theta^{-4}$, result in the jet closer to a tophat structure. 
The lack of gamma-ray radiation can be coherently explained within this framework, thus also implies that such events likely originate from off-axis-viewed GRBs.

\section{Summary and Conclusion}\label{sec:summary}

The recently discovered series of novel X-ray sources by EP have garnered widespread attention due to their uniquely long duration timescales and anomalously low peak energies. The multi-wavelength optical counterparts of these events (represented by EP240414a and EP241021a) often exhibit sudden rebrightening phenomena. Current theoretical models frequently discuss scenarios involving large-angle outflow afterglows 
(cocoon or structured jet wings,
~\citealp{Gianfagna_Piro_softXray_2025,
Yadav_Troja_Radioobservationspoint_2025}), 
refreshed shocks
~\citep{Busmann_OConnor_curiouscaseEP241021a_2025,
Srivastav_Chen_IdentificationOpticalCounterpart_2025}, 
multiple single-component jets 
(triple jets,
~\citealp{Shu_Yang_EP241021aMonthsduration_2025}; wobbling jets,
~\citealp{Gottlieb__LandscapeCollapsarOutflows_2025}), thermal-dominated cooling radiation
(cocoon/shocked ejecta,
~\citealp{vanDalen_Levan_EinsteinProbeTransient_2025,
Hamidani_Sato_EP240414aGammaRay_2025,
Sun_Li_fastXray_2025,
Zheng_Zhu_EP240414aaxisView_2025}), 
or merger/accretion induced collapse remnant interiors with magnetized wind nebula 
(also thermal-dominated,
~\citealp{Wu_Yu_EP241021acatastrophiccollapse/merger_2025}).

The aforementioned models have achieved their respective successes in interpreting the observational data from different perspectives. However, upon reviewing the historical observations and literature of typical GRBs, it is not uncommon to encounter similar late-time rebrightenings in the optical afterglow. 
In the past, these have always been explained as energy injections, refreshed shocks, relativistic wind bubbles, density jumps, and multi-component jets.
This paper does not intend to debate the merits and demerits of the above-mentioned models. Instead, we attempt to take a holistic view and attempt to establish a more universal model that is applicable across different redshifts, progenitor star properties, jet structures, and viewing angles. In this work, we demonstrate the possibility of such a more unified picture through three distinctive events.
A structured jet produced by the collapse of a massive star—potentially magnetized—is propagate through the wind-blown bubble surrounding the progenitor. The sudden rebrightening behavior of the afterglow occurs when the jet shocks pass through the terminal shock region of the stellar wind. This scenario provides a self-consistent explanation for the observed subclass of afterglows that exhibit rebrightening, particularly those characterized by a sharp rise and rapid decay. Our calculation results suggest a possible connection between off-axis-observed GRBs and some FXTs, and indicate that the large $\theta_{\rm{v}}/\theta_{\rm c}$ ratio may be a key to understanding such FXTs.

Considering potential variations in the wind strength and stochastic mass ejection processes of progenitor, the wind bubble may exhibit complex radial density structures. 
Under such conditions, the afterglow light curve could display even more intricate behaviors. 
We plan to conduct targeted numerical calculations to model these more complex scenarios.
In the future, we hope to further validate the reliability of this framework by delving deeper into historical data and utilizing more observational data from EP and SVOM missions.


\software{\texttt{Matplotlib}
	\citep{Hunter__Matplotlib2DGraphics_2007},
	\texttt{NumPy}
	\citep{Harris_Millman_ArrayprogrammingNumPy_2020},
	\texttt{emcee}
	\citep{ForemanMa_Hogg_emceeMCMCHammer_2013},
	\texttt{corner}
	\citep{ForemanMa__corner.pyScatterplotmatrices_2016},
	\texttt{Astropy}
	\citep{AstropyCo_Robitaill_Astropycommunitypython_2013},
	\texttt{SciPy}
	\citep{Virtanen_Gommers_SciPy1.0Fundamental_2020},
}

\acknowledgments
We thank the anonymous referee for the helpful comments and suggestions to improve this work.
J. R. is Supported by the Postdoctoral Fellowship Program and China Postdoctoral Science Foundation (grant No.~BX20250160), 
the Jiangsu Funding Program for Excellent Postdoctoral Talent (grant No.~2025ZB272), 
and the General Fund of the China Postdoctoral Science Foundation (grant No.~2024M763530).
D.M.W. is supported by the
Strategic Priority Research Program of the Chinese Academy
of Sciences (grant No.~XDB0550400), the National Key
Research and Development Program of China (No.~2024YFA1611704), and the National
Natural Science Foundation of China (NSFC; No.~12473049).
Z.G.D. is supported by the NSFC under No.~12393812.
Z.P.J. is supported by the NSFC under No.~12225305.
D.B.L. is supported by the NSFC under No.~12273005.
Y.W. is supported by the Jiangsu Funding Program for Excellent Postdoctoral Talent (grant No.~2024ZB110), 
the Postdoctoral Fellowship Program (grant No.~GZC20241916),
and the General Fund (grant No.~2024M763531) of the China Postdoctoral Science Foundation.	
L.L.Z. acknowledges the support by the Foundation of Guilin University of Technology (No. RD2500000384).
This work used data supplied by the UK Swift Science Data Centre at the University of Leicester and data obtained with Einstein Probe.

\clearpage
\restartappendixnumbering
\appendix
\section{Extended Figures}
\label{sec:appendix}

\begin{figure*}[htbp]
	\centering
	\includegraphics[width=0.9\textwidth]{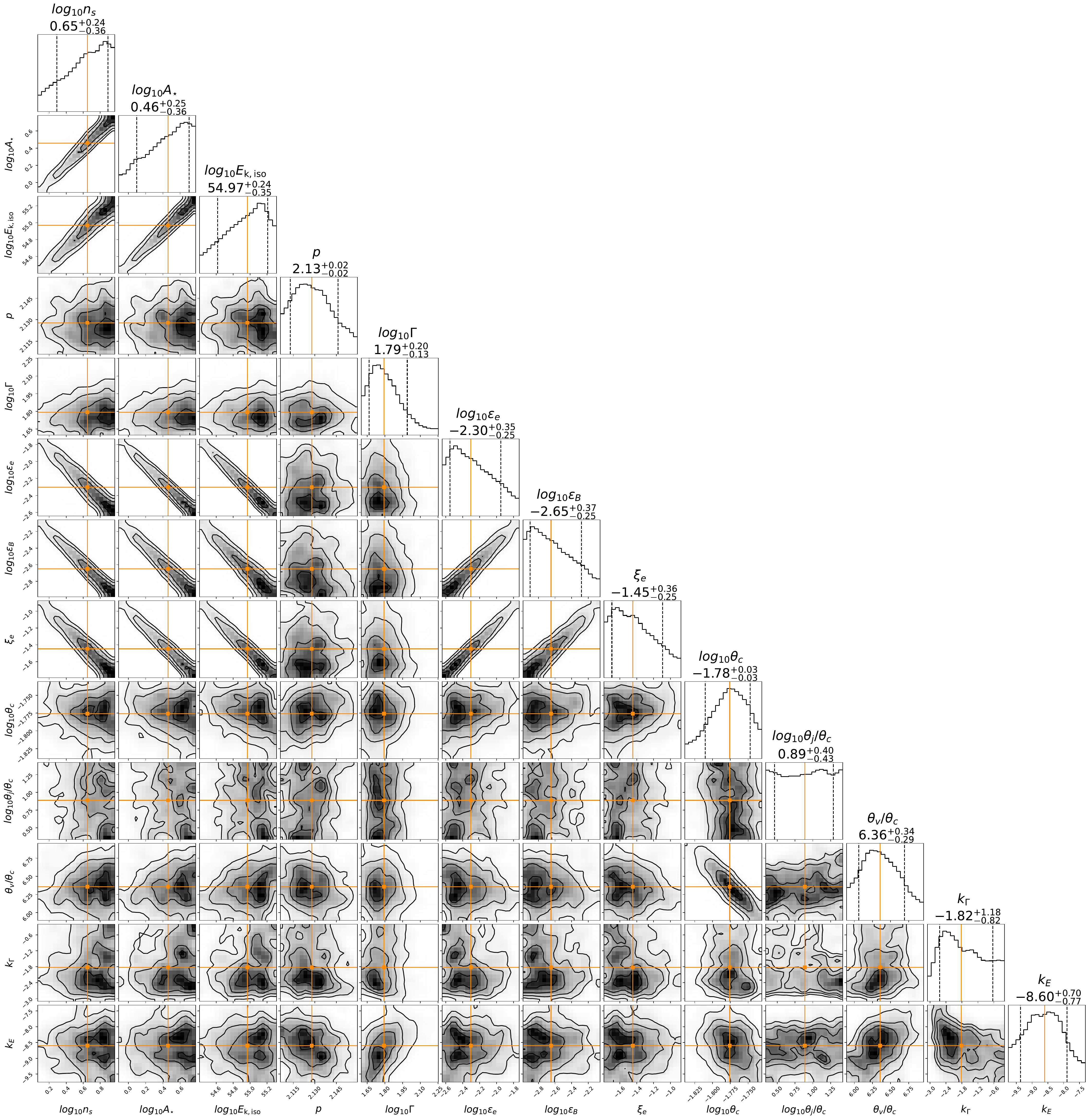}
	\caption{Corner plot of the MCMC posterior sample density distributions of EP240414a.}
	\label{corner_EP240414a}
\end{figure*}

\begin{figure*}[htbp]
	\centering
	\includegraphics[width=0.9\textwidth]{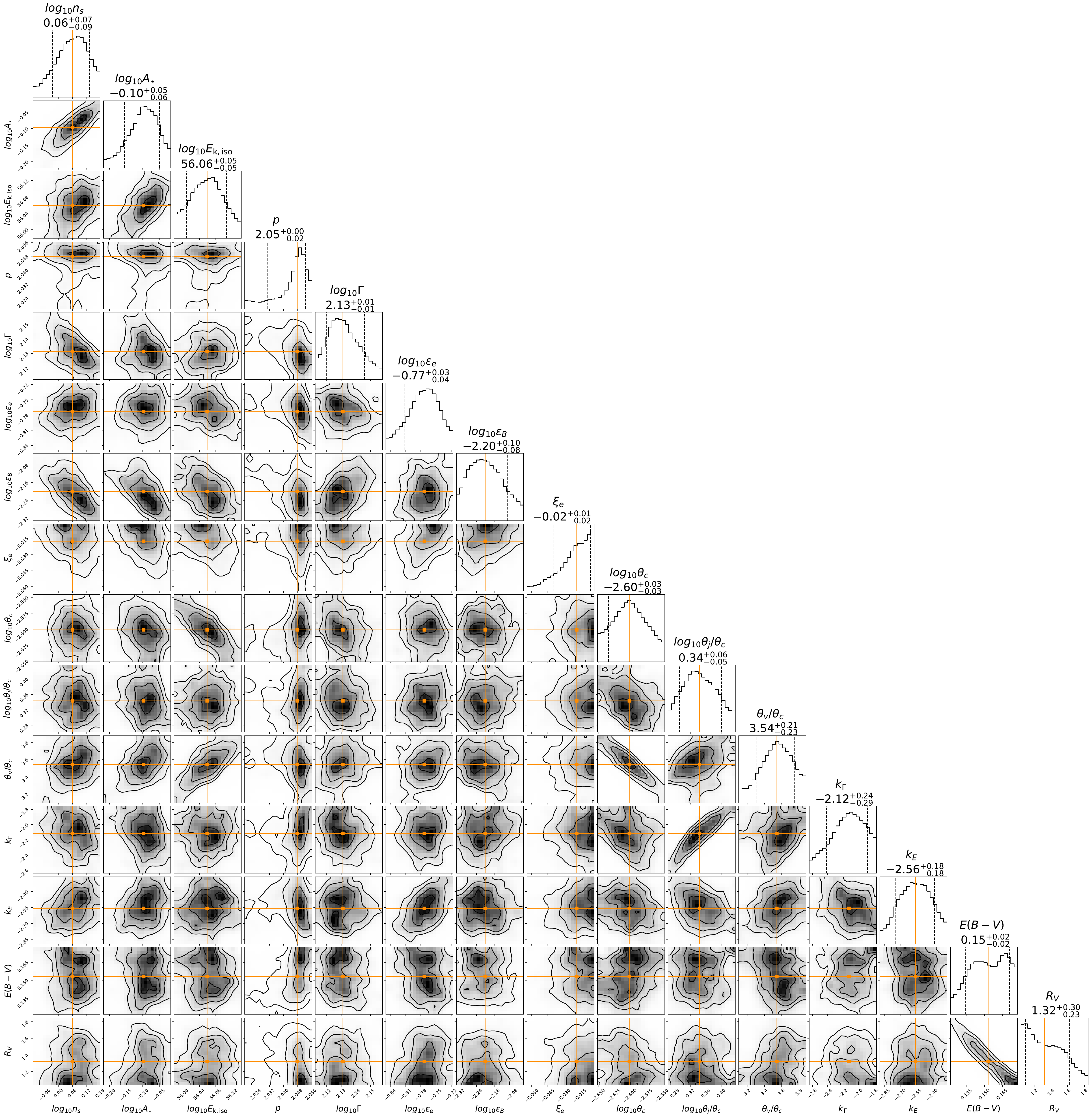}
	\caption{Corner plot of the MCMC posterior sample density distributions of GRB 240529A.}
	\label{corner_240529A}
\end{figure*}

\begin{figure*}[htbp]
	\centering
	\includegraphics[width=0.9\textwidth]{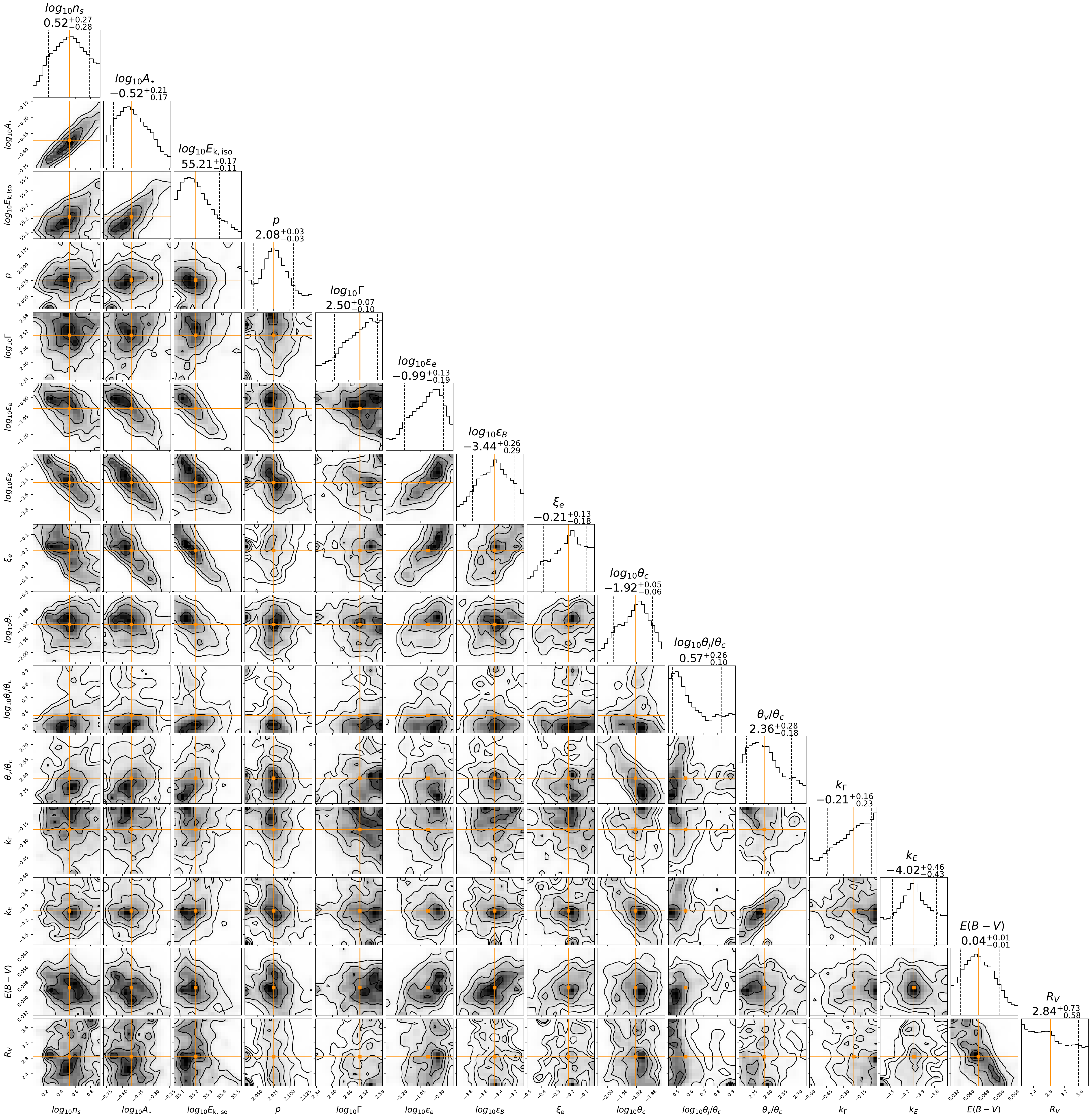}
	\caption{Corner plot of the MCMC posterior sample density distributions of GRB 240218A.}
	\label{corner_240218A}
\end{figure*}

\begin{figure}[htbp]
	\centering
	\includegraphics[width=0.8\textwidth]{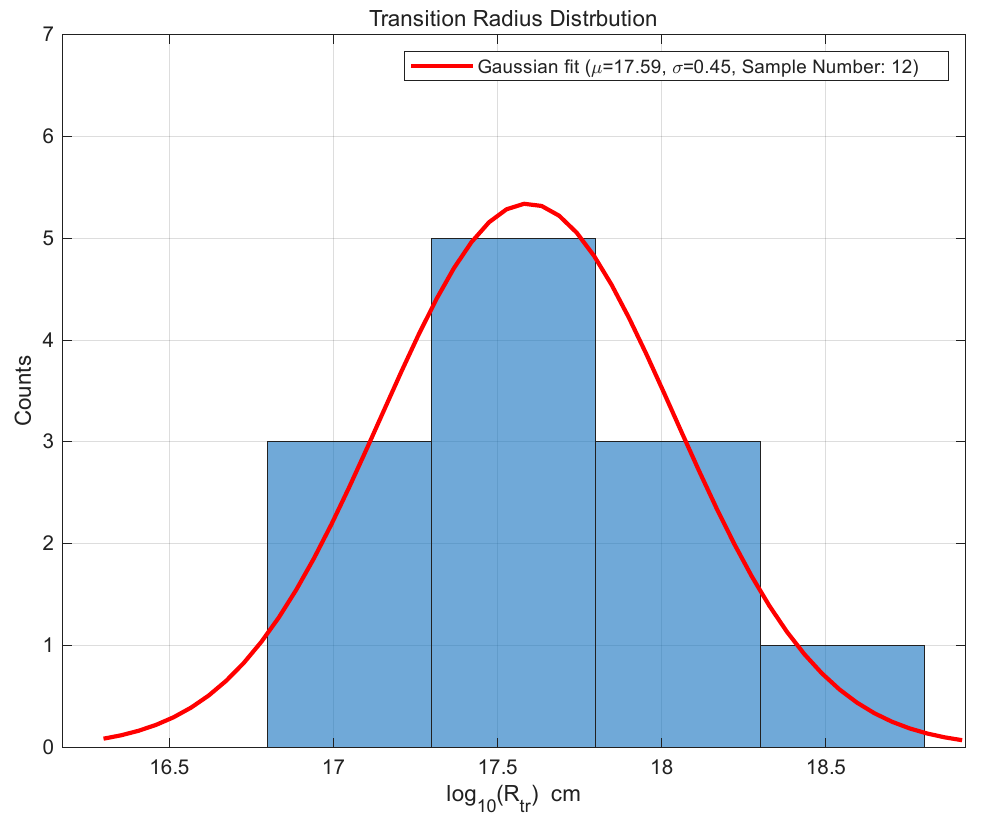}
	\caption{Transition radius distribution of samples, fitted by a lognormal function.}.
	\label{Radii}
\end{figure}

\clearpage


\bibliographystyle{aasjournal}
\bibliography{paper8.bib}
%

\end{document}